\documentclass[a4paper,english,aps,prb,twocolumn,floatfix,showpacs,amsfonts,amssymb,superscriptaddress]{revtex4} 
\usepackage[T1]{fontenc}
\usepackage[latin1]{inputenc}
\usepackage{graphics}
\usepackage{amssymb}
\usepackage{amsmath}
\usepackage{overpic}
\usepackage{ulem}

\newcommand{\be}{\begin{equation}}
\newcommand{\ee}{\end{equation}}
\newcommand{\bea}{\begin{eqnarray}}
\newcommand{\eea}{\end{eqnarray}}

\def\a{\alpha}
\def\b{\beta}
\def\e{\varepsilon}
\def\d{\delta}

\def\o{\omega}

\def\s{\sigma}

\def\D{\Delta}

\def\ra{\rightarrow}
\def\up{\uparrow}

\def\down{\downarrow}

\def\pd{\partial}

\def\bk{{\bf k}}

\def\bq{{\bf q}}

\def\bA{{\bf A}}

\def\bS{{\bf S}}

\def\nn{\nonumber}
\def\lb{\label}
\def\pref#1{(\ref{#1})}

\newcount\bozza \bozza=0
\ifnum\bozza=1
\newdimen\shift \shift=-2truecm
\def\lb#1{%
{\label{#1}\rlap{\kern\shift{$\scriptstyle#1$}}}}
\else\def\lb#1{\label{#1}} \fi

\begin{document}

\title{Current-current Fermi-liquid corrections to the 
  superconducting fluctuations on conductivity and diamagnetism}

\author{L.~Fanfarillo}

\affiliation{Department of Physics, Sapienza University of Rome, 
             P. le A. Moro 2, 00185 Rome, Italy}

\affiliation{Institute for Complex Systems (ISC), CNR, U.O.S. Sapienza, Sapienza University of Rome, 
             P. le A. Moro 2, 00185 Rome, Italy}

\author{L.~Benfatto}

\affiliation{Institute for Complex Systems (ISC), CNR, U.O.S. Sapienza, Sapienza University of Rome, 
             P. le A. Moro 2, 00185 Rome, Italy}

\affiliation{Department of Physics, Sapienza University of Rome, 
             P. le A. Moro 2, 00185 Rome, Italy}

\author{C.Castellani}

\affiliation{Department of Physics, Sapienza University of Rome, 
             P. le A. Moro 2, 00185 Rome, Italy}

\affiliation{Institute for Complex Systems (ISC), CNR, U.O.S. Sapienza, Sapienza University of Rome, 
             P. le A. Moro 2, 00185 Rome, Italy}

\date{\today}

\begin{abstract}
We analyze the behavior of the superconducting-fluctuations
contribution to diamagnetism and conductivity in a model system
having current-current interactions. We show that in proximity to a
Mott-insulating phase one recovers an overall suppression of the
fluctuating contribution to the conductivity with respect to
diamagnetism, in close analogy with recent experiments on the
underdoped phase of cuprate superconductors.
\end{abstract}


\maketitle

It is generally believed that, due to their low superfluid densities
and short correlation lengths, superconducting fluctuations (SCF) in
underdoped cuprates should be relevant for transport and thermodynamic
properties.  Such SCF have been widely highlighted 
in the pseudogap region by several experimental measurements, ranging from
diamagnetism\cite{Li_EPL_2005,Ong_PRB_2010,Cabo,Rigamonti} and 
Nernst effect\cite{Ong_Nernst, Ong_PRB_2010} 
to paraconductivity\cite{Leridon_PRB_2007,Rullier,Armitage_natphys11}. 
In particular, the survival of a large Nernst signal up to temperatures 
much larger than the superconducting transition temperature $T_c$ in the 
underdoped region  has been interpreted as the evidence 
for vortex-like phase fluctuations with a Kosterlitz-Thouless
(KT) character due to the quasi-two-dimensional (2D)
nature of the system\cite{Ong_PRB_2010,Li_EPL_2005}. At the same time, several 
authors\cite{Leridon_PRB_2007,Rullier} claimed that
paraconductivity in underdoped cuprates simply follows the $T$
dependence expected for the quasi-2D Aslamazov-Larkin (AL) regime
of gaussian Ginzburg-Landau (GL) SCF close to $T_c$ \cite{Varlamov_book}. 
This outcome motivated various investigations in the attempt to explain 
the experimental data on Nernst signal and diamagnetism within a GL-like
framework \cite{Ussishkin_PRL_2002, Levin_PRB_2004,Varlamov_Rapid_2011}. 

Regardless of the KT or GL character of fluctuations, there are two
outcomes of the experiments that are not expected for conventional
superconductors: 
(i) the range of temperatures where the fluctuation conductivity is
observed does not always match the one where a sizeable Nernst signal has been
reported\cite{Rullier,Armitage_natphys11}, 
and (ii) the SCF contribution to the conductivity is about two 
orders of magnitude {\it  smaller} than the fluctuating diamagnetism in the 
same system, as recently pointed out by Bilbro {\it et al.} in
Ref. \onlinecite{Armitage_2011}. To be more precise, we recall that in 2D the
contribution of SCF to the conductivity $\delta \sigma$
and diamagnetism $\delta\chi_d$ can be expressed both within the
GL\cite{Varlamov_book}  and
KT\cite{HN} theory in terms of the superconducting correlation length $\xi(T)$ as:
\be
\lb{para-dia}
\delta\chi_d = -\frac{k_BT}{\Phi^2_0 d}\xi^2(T) \ \ \ \ \ \ \ \ 
\delta\sigma = \frac{e^2}{16\hbar d} \frac{\xi^2(T)}{\xi^2_0}
\ee
Here $\Phi_0$ is the flux quantum, $\xi_0$ the low temperature correlation length 
and $d$ is the thickness of the effective 2D system (i.e. the interlayer
spacing for layered systems as cuprates).
Both $\delta\sigma$ and $\delta\chi_d$ diverge as $T$ approaches $T_c$ due to
the increase of the correlation length $\xi(T)$, with a $T$ dependence that is
power-law within the GL approach and exponential within the KT
theory. From Eq.s\ \pref{para-dia} one could express  
the ratio $\d\s/\d\chi_d$ as
\be
\lb{ratiodef}
\frac{\d\sigma}{|\delta \chi_d|}= \frac{\Phi_0^2 e^2 } {16 \hbar k_B T \xi^2_0 }
\frac{\xi^2_{\s}(T)}{\xi^2_{\chi_d}(T)}
\ee
where $\xi_{\s}$, $\xi_{\chi_d}$ are the correlations length extracted 
from paraconductivity and diamagnetism measurements, respectively.
Since one would expect that the same lenght scale is involved in both cases, 
$\xi^2_{\s}/\xi^2_{\chi_d}= 1$, rigth above $T_c$, $\d \s/\d\chi_d$  
should be of order $\sim 10^5  (\Omega\, \text{A/T})^{-1}$, while 
experimentally it turns out to be two orders of magnitude smaller than 
this\cite{Li_EPL_2005,Armitage_2011}
(see also discussion below Eq.\ \pref{ratio}).
Let us notice that the above discussion holds regardeless the nature of 
the SCF so that the the ratio between $\delta\sigma$ and $\delta\chi_d$ 
depends only on the properties of the system away from $T_c$.

In this paper we show that the quantitative disagreement between the
SCF contribution to diamagnetism and conductivity could be understood as
a consequence of current-current interactions in a doped Mott
insulator. 
The possible relevance of this kind of interactions to the physics of cuprates has
been suggested within several contexts, ranging from the gauge-theory
formulation for the $t-J$ model\cite{Lee_2006} to the theoretical
approaches emphasizing the role of microscopic
currents\cite{Varma, Chakravarty}.  As a paradigmatic example we focus
on the $t-J$ model within the slave-boson approach, where
the Hartree-Fock (HF) correction of
the quasiparticle dispersion leads to a difference 
between the quasiparticle current and velocity in the usual
Landau Fermi-liquid (FL) language \cite{Nozieres}. The effect of this
dichotomy on the GL functional for SCF can be accounted for within 
a general field-theory for the t-J model\cite{Lee_2006,Ng_PRB}, which
includes fluctuations both in the
particle-particle (p-p) and particle-hole (p-h) channel. 
By computing the AL contribution to diamagnetism and conductivity we find 
that the role of Landau FL corrections differs in the
static or dynamic limit. As a consequence the SCF contribution to
diamagnetism and conductivity scales with a different prefactor,
leading to the suppression of paraconductivity with respect to
fluctuation diamagnetism when the Mott insulator is approached, in
analogy with experiments in cuprates.

Let us start from the slave-boson version of the $t-J$ model, 
\be
\lb{htj}
H = -t \d \sum_{\left\langle i,j \right\rangle \, \sigma}(c^{\dagger}_{i
  \sigma}c_{j \sigma} + H.c.)+ J  \sum_{\left\langle i,j
  \right\rangle}  \bS_i \bS_j
\ee
where $t$ is the electron hopping, $J$ is the exchange interaction
between electron spins,  $c^{\dagger}_{i \sigma} (c_{i \sigma})$ is
the fermionic creation (annihilation) operator and
$\bS_i=\Psi^{\dagger}_i \frac{\boldsymbol{\sigma}}{2}
\Psi_i$ is the spin operator, with $\Psi_i= (\begin{tiny} \begin{matrix} c_{i \up} 
\\ c_{i \down} \end{matrix} \end{tiny})$. 
The sum is extended over all the $\left\langle i,j
\right\rangle$ nearest-neighbor pairs on a square lattice, and we use units such that 
the lattice spacing and $\hbar=c=e=1$. In
Eq.\ \pref{htj} the decomposition of the electron operator in a
fermionic spinon and bosonic holon part has been already carried
out\cite{Lee_2006}, and only the fermionic degrees of freedom have been
retained. Here we neglect the boson fluctuations, and we
assume that slave bosons are always condensed, leading to the
suppression factor $\delta=2x/(1+x)$ of the hopping, scaling with the
doping $x$. 
The interaction term of the Eq.\pref{htj} contains contributions both from the 
p-p and the p-h channel. We introduce the operators 
\bea
\lb{dop}
\Phi^c_\a(\bq) &=& \sum_{\bk \s} \cos \bk_\a  \  c^{\dagger}_{\bk+\bq/2, \s}  c_{\bk-\bq/2, \s}, \\
\lb{cop}
\Phi^s_\a(\bq) &=& \sum_{\bk \s} \sin \bk_\a  \ c^{\dagger}_{\bk+\bq/2, \s}  c_{\bk-\bq/2, \s}, \\
\lb{pop}
\Phi^{\Delta}(\bq) &=& \sum_\bk \gamma_d(\bk) \ c_{-\bk+\bq/2,\down}  c_{\bk+\bq/2, \up} 
\eea
where $\a=x,y$ and 
$\gamma_d (\bk) = \cos \bk_x - \cos \bk_y$, so the interaction term reads
\be\lb{hint}
\sum_{\left\langle i,j \right\rangle}  \bS_i \bS_j = - g\sum_{\bq}
\bigg(\frac{1}{2} \sum_{\alpha} \Phi_{\a \, \bq}^c\Phi_{\a \, \bq}^c 
+\Phi^s_{\a \, \bq} \Phi^s_{\a \, \bq}\bigg) 
+ {{\Phi_\bq^{\Delta}}^*}\Phi^{\Delta}_\bq
\ee
with $g=3J/4$. The RHS of Eq.\ \pref{hint} represents
the interaction in the p-h density ($\Phi^c$), p-h current
($\Phi^s$) and p-p ($\Phi^\Delta$) channel, respectively\cite{overcounting}.

We decouple\ \pref{hint} by means of the Hubbard-Stratonovich (HS) 
trasformation both in the p-p and in the p-h channel. 
After the integration of the fermions the action reads:
\be\lb{S_Eff}
S = \sum_{q} \bigg(\sum_{\a}\frac{|\phi^c_{\a \, q}|^2}{2 g} + 
\frac{|\phi^s_{\a \, q} |^2}{2g}\bigg)+\frac{|\Delta_q|^2}{g} 
 - \mathrm {Tr}\log \hat A_{kk'}.
\ee
Here $\phi^c$, $\phi^s$ and $\Delta$ are the HS fields, the trace acts over 
momenta, frequencies and spins, $k\equiv(\bk,i\e_n)$, $q\equiv(\bq,i\omega_m)$, 
and $\e_n$, $\omega_m$ are the Matsubara fermion and boson frequencies,
respectively. The $\hat{A}_{k,k'}$ matrix is defined (in the usual Nambu
notation) as
\bea
\hat{A}_{k' k}&=& -\big[i\omega_n +  
\sin(
\scriptstyle \frac{\bk +\bk'}{2}
\displaystyle )_\a \phi^s_{\a \ k -k'}\big] \hat{\tau}_0 \nn \\
\lb{akk}
&+& \bigg[\xi^0_{\bk} - \cos(
\scriptstyle \frac{\bk +\bk'}{2}
\displaystyle)_\a \phi^c_{\a \ k -k'}\bigg] \hat{\tau}_3  
- \bigg[\Delta_{k - k'} \gamma_d(
\scriptstyle\frac{\bk +\bk'}{2}
\displaystyle) \bigg] \hat{\tau}_1 
\eea
where $\xi^0_\bk$ is the bare dispersion
\be\lb{xi0}
\xi^0_\bk=-2t\delta  (\cos \bk_x + \cos \bk_y) -\mu
\ee
The Eq.\ \pref{akk}  can be decomposed as 
$\hat{A}_{k,k'} = - \hat{G}_0^{-1}\delta_{k,k'}+\hat{\Sigma}_{k-k'}$. 
Here $\hat{G}_0^{-1}\delta_{k,k'}$ contains the $q=0$ saddle-point
values ($\phi_0^c, \phi_0^s,\Delta_0$)
of the HS fields obtained by minimization of the mean-field action
\be
\lb{smf}
S_{MF} =  \sum_{q} \bigg(\sum_{\a}\frac{|\phi^c_{\a \, q}|^2}{2 g} + 
\frac{|\phi^s_{\a \, q} |^2}{2g}\bigg)+\frac{|\Delta_q|^2}{g} 
-\mathrm {Tr}\log (\hat{G}_0^{-1}),
\ee
while $\hat{\Sigma}_{k-k'}$ contains the fluctuating parts of the HS fields. 
The standard GL functional above $T_c$ will then be given by the
expansion of Eq.\ \pref{S_Eff} around the mean-field action \pref{smf} as
$S_{GL}	=  \sum_n \mathrm {Tr}[\hat{G}_0\hat{\Sigma}]^n/n$, with 
$\Delta_0=0$. As far as the
saddle-point values in the p-h channels are concerned, one can easily
see that $\phi_0^s=0$, while $\phi^c_{0} \neq 0$ satisfies the following 
self-consistent equation: 
\be\lb{phic}
\phi^c_0 = \frac{2g}{N} \sum_\bk \cos  \bk_\a f(\beta \xi_\bk)
\ee
where $f(x)$ is the Fermi function, $\beta=1/T$  and $\xi_\bk$ is the quasiparticle
dispersion, given by the $\hat \tau_3$ term of Eq.\
\pref{akk}:
\be 
\lb{xiqp}
\xi_\bk=-(2t\delta+\phi_0^c)  (\cos \bk_x + \cos \bk_y)-\mu
\ee
As one can see, the $\phi_0^c$ value corresponds thus to the standard HF 
correction of the quasiparticle dispersion. 

In order to compute the contribution $\delta\chi(q)$ of SCF to the electromagnetic
response function we need to introduce in the effective action also the 
electromagnetic potential $\bA$. 
For the model \pref{htj} this can be done via the Pierls
substitution $c^\dagger_ic_{i+\alpha}\rightarrow
c^\dagger_ic_{i+\a}e^{-i  A^\a_i}$. which modifies the 
$\hat{A}_{k k'}$ matrix with two additional contributions
\bea
\lb{akkbis1}
&-& 2 t \d \,  \sin(
\scriptstyle \frac{\bk +\bk'}{2}
\displaystyle )_\a \ A^\a_{k' - k} \  \hat {\tau}_0 \\
\lb{akkbis2}
&+& t \d \,  \cos(
\scriptstyle \frac{\bk +\bk'}{2}
\displaystyle )_\a  A^\a_{k' - k + s} A^\a_{-s} \ \hat{\tau}_3.
\eea 
Eq.\ \pref{akkbis1} corresponds to the usual term $-{\bf J}\cdot \bA$,
where 
\be
\lb{defj}
J_\a\equiv \pd_{\bk_\a} \xi^0_\bk =2t\delta\sin \bk_\a
\ee
is the quasiparticle current. Notice that in the presence
of HF corrections to the quasiparticle dispersion \pref{xiqp} the quasiparticle
current $J_\alpha$ is different from the quasiparticle velocity 
\be
\lb{defv}
v_\a\equiv \pd_{\bk_\a} \xi_\bk=(2t\delta+\phi_0^c)\sin \bk_\a
\ee
In the usual Landau FL language, these two quantities are related by the Landau FL 
$F_1^s$ corrections as $J_\a= v_\a(1+ F_1^s/3)$ for an isotropic system in 
three dimensions\cite{Nozieres}. 
As we shall see, in our approach the analogous role of Landau FL corrections 
will be played by the HS fields related to p-h fluctuations. 
Neglecting the diamagnetic contributions, that are not relevant for the following 
discussion, the leading terms in $\bA$ of the action $S_{GL}(\bA,\Upsilon^{HS})$ 
($\Upsilon^{HS}=\phi^c,\phi^s,\Delta$) can be written in a compact form as:
\be\lb{sflA}
S_{GL}(\bA,\Upsilon^{HS})= 
-\frac{1}{2} \chi^{\a \b}_{j j} \ A^\a_{q} A^\b_{-q} +  A^\a_{q} F^\a(\Upsilon^{HS})
\ee
where $\chi^{\a \b}_{j j}$ is the mean-field current-current correlation function
\be
\lb{defxjj}
\chi^{\a \b}_{j j}(q)= -\frac{2}{N} \sum_{\bk} (\pd_{\bk_\a} \xi_0) (\pd_{\bk_\b} \xi_0)  
\frac{f(\b \xi_{\bk+\frac{\bq}{2}})-f(\b \xi_{\bk-\frac{\bq}{2}}) }{i\omega_n + 
\xi_{\bk+\frac{\bq}{2}} - \xi_{\bk-\frac{\bq}{2}}} 
\ee
and $F^\a(\Upsilon^{HS})$ is a function of the $\Upsilon^{HS}$ fields which
describes the connection of the electromagnetic potential to the HS
fields.  The current-current response function $\Lambda^{\alpha \beta}(q)$ 
can be computed from the partition function
$Z[\bA] = \int{\cal D} \Upsilon^{HS} e^{-S_{GL}(\bA, \Upsilon^{HS})}$
of the model as
\be
\lb{corr}
\Lambda^{\alpha \beta}(q)= \left .\frac{\delta \ln{Z[\bA]}}
{\d A^\a_q\d A^\b_{-q}}\right|_{A^\a_q=A^\b_{-q}=0}
=\chi_{jj}^{\a \b}(q) +\delta \chi_{jj}^{\a\b}(q)
\ee
where $\chi_{jj}^{\a \b}(q)$ simply follows from the quadratic 
term of Eq. \pref{sflA}, while $\d\chi_{jj}^{\a \b}(q)$ is the contribution coming from the fluctuating 
modes coupled to $\bA$, and depends on the explicit form of the $F(\Upsilon^{HS})$ function:
\be
\lb{deltachi}
\delta \chi_{jj}^{\a\b}(q)= \langle  F^{\a}(\Upsilon^{HS}) F^{\b}(\Upsilon^{HS})\rangle
\ee 
Starting from the current-current response $\Lambda^{\a \b}(q)$ \pref{corr}
the paraconductivity simply follows from the dynamic limit
$(\bq=0,\o\ra 0)$  after analytical continuation of the Matsubara frequency 
$\o_m$ to the real frequency $\o$

\be
\lb{sfdc}
\delta\sigma=[{\mathrm{Im}}\,\delta\chi_{jj}^{\a\a}(\o,\bq=0)/\o]_{\o\ra
  0},
\ee
while the fluctuations contribution to the diamagnetism is
connected instead to the static limit $(\bq\ra 0,\o=0)$, 
\be
\lb{sfdia}
\delta \chi_d= - \left[\delta \chi_{jj}^{\, t}(\bq,\o=0)/\bq^2\right]_{\bq\ra
  0}.
\ee
where $\d \chi_{jj}^{\, t}$ is the transverse part of the fluctuation 
correction to the current-current response function.
In the absence of fluctuations in the p-h channel, only the pairing
field is coupled to $\bA$ so that ${\bf F}(\Delta)\sim {\bf p} \Delta^2_p$ and
one recovers the standard AL correction (see Eq. \pref{nosfcorr}
below). In our case, one immediately sees from Eq.\ \pref{akk} and
\pref{akkbis1} that the $\phi_\a^s$ field appears in the actions with
the same structure of the electromagnetic potential $A^\a$, i.e. it is
coupled to the fermionic current. Thus, we expect that it will contribute 
to the $\delta\chi_{jj}^{\a\b}(q)$ correction \pref{corr} above.

\begin{figure}[thb]
\includegraphics[scale=0.55,clip=true]{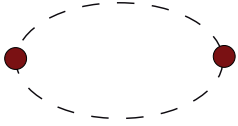}
\vspace{0.3cm}
\caption{(color online) AL contribution of the SCF to the current-current correlation
  function. The dashed lines represent the SCF propagator, while the
  red dots reduce to the constant $c$ in the ordinary
  case Eq.\ \pref{nosfcorr}, and to the momentum and frequency
  dependent vertex $c(q)$ in the presence of current-current interactions, see Eq.\
  \pref{sfcorr}. }
\label{fig-curr}
\end{figure}

With lengthy but straightforward calculations one obtains that the
effective action \pref{sflA} at leading order in the gauge and HS
fields is given by
\begin{widetext}
\bea
S_{GL}(\bA,\phi^s,\Delta)&=& -\frac{1}{2} \chi_{jj}^{\a \b}  A^\a_{q} A^\b_{-q} - 
\frac{1}{2} \frac{\chi_{j j}^{\a \b}}{2t\d}  \big( A^\a_{q} \phi_{-q}^{s\,\b} + 
\phi_q^{s \,\a} A_{-q}^{\b}\big) + \frac{1}{2}\Bigg[g^{-1}- 
\frac{\chi_{j j}^{\a \b}}{(2 t\d)^2}\Bigg]\phi_q^{s \,\a} \phi_{-q}^{s \, \b} + \nn \\
\lb{S_EffA}
&+&\Bigg[g^{-1} - \Pi(\omega) +  c \bq^2\Bigg]\Delta^{*}_{q} \Delta_q
-  c' \bq \bigg(\bA _{q'} +  \frac{{\boldsymbol \phi}^s _{q'}}{2t\d } \bigg) 
\big( \Delta^{*}_{q} \Delta_{q-q'}+ \Delta^{*}_{q+q'}\Delta_{q}\big) 
\eea
\end{widetext}
where summation over repeated indexes is implicit and only the terms relevant 
for the following discussion are included.
\begin{figure}[thb]
\includegraphics[scale=0.55,clip=true]{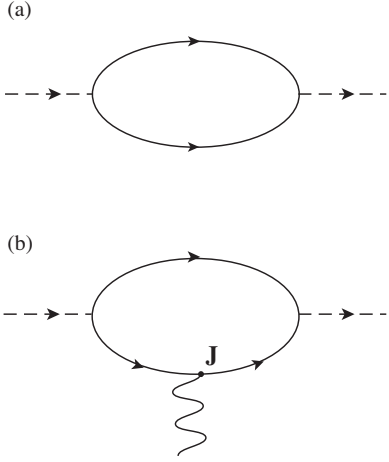}
\vspace{0.3cm}
\caption{Fermionic bubbles: (a) is the $\Pi(q)$ bubble associated to the $\Delta^2$ term, 
while (b) is the fermionic bubble relative both to $\bA\Delta^2$ and $\phi^s\Delta^2$ terms. 
The solid lines are the Green's functions ${\cal G}$ containing the full 
quasiparticle dispersion $\xi$. The dashed lines are the pairing field $\Delta$. 
The wavy line represents either the electromagnetic potential $\bA$ or the current 
field $\boldsymbol{\phi}^s$, both associated to the quasiparticle current ${\bf J} \propto \pd \xi^0$.}
\label{bubble}
\end{figure}
The $c$ and $c'$ terms in Eq.\ \pref{S_EffA} follow from the expansion to leading order
in $\bq$ of the fermionic bubbles associated to the $\Delta^2$ term and to the 
$\bA\Delta^2$ and $\phi^s\Delta^2$ terms, respectively (see Fig.\ \ref{bubble}). 
In the former case for example one
expands
\be
\Pi(q) = -\frac{T}{N} \sum_{k} \gamma_d^2(\bk) {\cal G}_{k+q/2}{\cal G}_{-k+q/2} 
\simeq \Pi(\omega) - c\bq^2 . 
\ee
where the quasiparticle Green's function, ${\cal G}_{k} = (i \o_n - \xi_{\bk})^{-1}$, contains the full
dispersion $\xi_\bk$. As a consequence, 
$c$ is proportional to the second-order derivative of  $\cal G$, 
which in turn scales as $(\pd \xi_{\bk_\a})^2\equiv v_\a^2$.
In the case of the $c' \bq$ term instead one carries out a single
derivative of the fermionic bubble which contains already a current 
insertion ${\bf J}$, associated to each $\bA$ or $\boldsymbol{\phi}^s$ field,
see Eq.s\ \pref{akk} and \pref{akkbis1} and Fig.\ \ref{bubble}. Thus, 
we have in a short-hand notation:
\be
c\propto (\pd \xi_{\bk_\a})^2 \sim v^2_F , \ \ \ \ \ \ 
c'\propto \pd\xi_{\bk_\a} \pd\xi^0_{\bk_\a}\sim  v_F J_F
\ee
where $v_F$, $J_F$ are the average values on the Fermi surface.
If we do not consider the interactions in the p-h channel leading to
the HF correction of the band dispersion we would get $\xi_\bk=\xi^0_\bk$ 
(see Eq.\ \pref{xiqp}) and the velocity coincides with the current, so that $c=c'$.
In this case, as mentioned above, the electromagnetic potential $\bA$  is 
coupled only with the pairing field $\Delta$ via a term $\sim
c \bA {\bf p} \Delta_p^2$, so that the fluctuation correction 
to the current-current response function \pref{corr} is given by the usual 
AL contribution\cite{Varlamov_book} (see Fig.\ref{fig-curr}):
\be 
\lb{nosfcorr}
\delta\chi^{\a\b}(q)=  c^2  T   \sum_{p}(2 {\bf p} + \bq)_\a (2 {\bf p} + \bq)_\b
L(p)L(p+q)
\ee 
where $L(p)=\langle\Delta^*_p \Delta_p\rangle$ is the propagator 
of the SCF. 
Since $c\propto v_F^2$ is constant, 
the SCF contribution to  the diamagnetism \pref{sfdia} and
conductivity \pref{sfdc} is in both cases proportional to $v_F^4$, and the ratio 
$\d \s/\d \chi_d$ is expected to be of order ${\cal O}(1)$.
Such a result changes in the presence of p-h interactions.
The elimination via gaussian integration of the current field $\phi^s$ from Eq. \pref{S_EffA} leads to 
\bea
S_{GL}(\bA,\Delta)&=& -\frac{1}{2} \chi_{jj}^{\a \b} \big(1 + Z(q)\big)  A^\a_{q} A^\b_{-q} + \nn \\
&+&\big(g^{-1} - \Pi(\omega) +  c \bq^2\big)\Delta^{*}_{q} \Delta_{q} + \nn \\
\lb{S_EffAbis}
&-&  c(q) \bq \bA _{q'} \big( \Delta^{*}_{q} \Delta_{q-q'}+ \Delta^{*}_{q+q'}\Delta_{q}\big) 
\eea
where 
\be\lb{LFLcorr}
Z(q)=\frac{\chi_{j j}(q)/(2t\d)^2}{1/ g- \chi_{j j}(q)/(2t\d)^2}
\ee
and
\be\lb{c}
c(q) = c' (1+Z(q)).
\ee
We notice that in Eq.s\ \pref{S_EffAbis}-\pref{c} we used a simplified notation valid in the 
$q \ra 0$ limit relevant for the following discussion, so that the current-current 
correlation function $\chi_{jj}$ refers to its diagonal part $\chi_{jj}^{\a\a}$ only.
At finite $q$ the above expressions must be properly extended to account for the full 
matrix structure of the response functions.   
From the definition \pref{S_EffAbis} and using Eq.\ \pref{corr} we compute the SCF 
contribution to the current-current response function. We find a
correction to the mean-field value current-current response function \pref{defxjj} that
leads to the RPA resummation of $\chi_{jj}$ as
\be
\chi_{jj}^{RPA}(q) = \chi_{jj}(q) + Z(q) \chi_{jj}(q) =  
\frac{\chi_{j j}(q)}{1- g \chi_{j j}(q)/(2t\d)^2},
\ee
and a second term that represents the SCF contribution, which
generalizes the standard AL result \pref{nosfcorr} with a momentum and
frequency dependent vertex $c(q)$:
\be 
\lb{sfcorr}
\delta\chi^{\a\b}(q)= T \sum_{p}c(q)^2(2 {\bf p} + \bq)_\a (2 {\bf p} + \bq)_\b
L(p)L(p+q).
\ee 
From Eq.\ \pref{defxjj} one can easily see that in the dynamic limit 
$\chi_{jj}=0$, so that $Z=0$ in Eq.\ \pref{LFLcorr} and from Eq.\ \pref{c}
$c(q)=c'$. In the opposite static limit instead we can rewrite
$\chi_{jj}$ in Eq.\ \pref{defxjj} as
\bea 
\chi_{jj} &=& -\frac{2}{N} \sum_{\bk} (\pd_{\bk_\a} \xi^0_\bk)^2  \pd_{\xi}f(\b \xi_\bk) \nn \\
&=&\frac{2}{N} \sum_k \frac{J_\a}{v_\a}(\pd^2_{\bk_\a} \xi^0_\bk) f(\b \xi_\bk)
\eea
By direct comparison with the self-consistent equation \pref{phic} for
the density field one immediately sees that $\chi_{jj}=2 t
\d(J_F/v_F)(\phi_0^c/g)$. From Eq.\ \pref{LFLcorr} it then follows
that $1+Z=v_F/J_F$, leading to $c(q)=c$ in Eq.\ \pref{c}. We then find
that the vertices relative to the SCF contribution to the conductivity
and diamagnetism are quantitatively different:
\bea
\lb{cdyn}
c(q) &\propto& v_F J_F \sim (2t\d+\phi^c_0) 2\delta t  \quad (\bq=0, \o\ra 0) \\
\lb{cstat}
c(q) &\propto& v_F^2  \sim  (2t\d+\phi^c_0)^2   \quad (\bq\ra 0,
\o= 0). 
\eea
The difference in the two limits reflects in a difference in the overall
prefactors of the SCF contribution to the conductivity and
diamagnetism. Indeed, from Eqs.\ \pref{sfdc}-\pref{sfdia} one has that 
\be
\lb{ratio}
\frac{\d\sigma}{|\delta \chi_d|}  \propto 
\frac{((2 t\d+\phi^c_0) 2 t\d)^2}{(2 t\d+\phi^c_0)^4} 
\propto\delta^2
\ee
i.e. the fluctuation conductivity is suppressed by the proximity to
the Mott insulator by a factor that depends on the doping. 

We notice that in the presence of HF corrections the gauge-invariant
form of the GL functional for SCF is recovered in a non trivial way. Indeed, in the usual case
the coupling to the gauge field $\bA$ can be obtained by the
minimal-coupling substitution $\bq\ra\bq-2 \bA$ in the $\bq^2\Delta^2$
term of the Gaussian propagator. This leads immediately
to the term linear in $\bA$, $c \, \bA\cdot\bq \Delta^2$, needed to compute the AL
correction, see Eq.s\ \pref{sflA}-\pref{deltachi}. In our case by direct
inspection of Eq.\ \pref{S_EffA} one finds instead two different
coefficients $c,c'$ in the $\bq^2\Delta^2$ and $\bA\cdot\bq \D^2$
terms, as we explained above. However, by integrating
out the $\phi^s$ field the coupling of SCF to the
gauge field is described in general by a term $c(q)\bA\cdot\bq \D^2$, (see Eq.\ \pref{S_EffAbis}),
where $c(q)$ is given by Eq.\ \pref{c}. As a consequence, one recovers
the minimal-coupling prescription only in the static limit
\pref{cstat} where $c(q) = c$ \cite{gaugeinvariance}.

According to the discussion below Eq.\ \pref{para-dia} above, the result \pref{ratio} 
can be recast as an estimate of the ratio between the correlation lengths extracted  
experimentally from paraconductivity and diamagnetism.
For example using data reported in Ref.\ [\onlinecite{Armitage_2011}] one
has that at $T \sim 24$\,K, $\d\s \sim 10^5 (\Omega\, \text{m})^{-1}$, and 
$\d\chi_d \sim 60\, \text{A/mT}$. 
Using in Eq.\ \pref{para-dia} $\xi_0\sim 1$\,nm 
and $d\sim 15$\, \AA \ as appropriates for cuprates, we obtain that
$k_BT/d\Phi_0^2=0.053\, \text{A/mT}$ and $e^2/16\hbar d \xi_0^2=10^4 (\Omega\, \text{m})^{-1}$, so that 
$\d\s/\d\chi_d\propto \xi^2_{\s}/\xi^2_{\chi_d}\sim 10^{-2}$. Such a strong
suppression of the paraconductivity with respect to diamagnetism, that has
been rephrased in  Ref.\ [\onlinecite{Armitage_2011}] in terms of the vortex
diffusion constant valid only in the case of KT fluctuations, 
can be more generally attributed 
from Eq.\ \pref{ratio} to the overall $\delta^2$ factor due to
the proximity to the Mott-insulating phase. 
We checked that the estimate of the ratio \pref{ratio} within the mean-field 
solution of the $t-J$ model is quantitatively larger by a factor ten than the one 
experimentally found, since a prefactor $\sim 2t/\phi^c_0$ partly compensate 
the $\d^2$ suppression. Such a quantitative disagreement is
reminiscent of analogous limitations of the mean-field approach already
discussed in the literature in the contexts of other physical
quantities, as for example the scaling of the superfluid-density
depletion with doping\cite{Ng_PRB}. 
Moreover, at low doping\cite{Kotliar,Lee_2006,Ng_PRB} one should also include 
the boson fluctuations neglected so far, which give a temperature $T_B$ for the
boson condensation smaller than the one for the gap opening,
leading to a suppression of the critical temperature $T_c$ with
respect to its mean-field value $T_{MF}$. In such a regime, the static limit of the $1+Z$
correction in Eq.\ \pref{c} above could not be simply given by $v_F/J_F$: 
nonetheless, the difference between the static and dynamic
limit still holds, possibly leading only to a quantitative difference with respect
to the result \pref{ratio}. Finally, we notice that the experimental
estimate of the SCF contribution to paraconductivity reported in Ref.\
[\onlinecite{Armitage_2011}] is based on the scaling of the 
conductivity at frequencies large enough with respect to 
quasiparticle dissipation. As a
consequence, it is worth making a comparison with the present results,
that have been derived in the clean limit. 

In summary, we analyzed the SCF contribution to conductivity and diamagnetism in
the presence of current-current interactions, by using as a
paradigmatic example the slave-boson formulation for the $t-J$ model.
By explicitly constructing the GL fluctuation functional in the
presence of HF corrections we showed that  current-current interactions, 
 needed to recover the gauge-invariant form of the GL functional,
modify the transport coefficients
leading to a momentum and frequency dependence of the vertex
$c(q)$ entering the AL expression for the SCF contribution.  
Since different limits are involved in the definition of 
paraconductivity and diamagnetism, we obtain a different prefactor in the
two cases, with a suppression of the paraconductivity because of the proximity to 
the Mott-insulating phase, as recently shown experimentally in
cuprates\cite{Armitage_2011}. Even though a mean-field approach to the $t-J$
model is not satisfactory from the quantitative point of view, the
different strength of SCF contribution to conductivity and
diamagnetism is more general, since it is a consequence of the
existence of a sizeble difference between quasiparticle current and velocity. A  
quantitative comparison with experiments remains an interesting
theoretical challenge, that certainly deserves further investigation.

\end{document}